\documentclass[12pt]{article}
\usepackage{amsmath,amssymb,bm,graphicx}
\usepackage{color} 

\newcommand{\delslash}{\not \! \partial}

\begin{document}


\begin{center}
{\Large{\bf Operatorial characterization of Majorana neutrinos}}
\end{center}
\vskip .5 truecm
\begin{center}
{\bf { Kazuo Fujikawa
}}
\end{center}

\begin{center}
\vspace*{0.4cm} 
{\it {
Interdisciplinary Theoretical and Mathematical Sciences Program,\\
RIKEN, Wako 351-0198, Japan
}}
\end{center}
\vspace*{0.2cm} 
\begin{abstract}
The Majorana neutrino $\psi_{M}(x)$ when constructed as a superposition of chiral fermions such as $\nu_{L} + C\overline{\nu_{L}}^{T}$ is characterized by  $ ({\cal C}{\cal P}) \psi_{M}(x)({\cal C}{\cal P})^{\dagger} =i\gamma^{0}\psi_{M}(t,-\vec{x})$, and the CP symmetry describes the entire physics contents of Majorana neutrinos.  Further specifications of C and P separately could lead to difficulties  depending on the choice of C and P. 
The conventional $ {\cal C} \psi_{M}(x) {\cal C}^{\dagger} = \psi_{M}(x)$  with well-defined P 
is naturally defined when one constructs the Majorana neutrino from the Dirac-type fermion.
In the seesaw model of Type I or Type I+II where the same number of left- and right-handed  chiral fermions appear, it is possible to use the generalized Pauli-Gursey transformation to rewrite the seesaw Lagrangian in terms of  Dirac-type fermions only; the conventional C symmetry then works to define Majorana neutrinos.   In contrast, the ``pseudo C-symmetry'' $\nu_{L,R}(x)\rightarrow C\overline{\nu_{L,R}(x)}^{T}$  (and associated ``pseudo P-symmetry''), that has been often used in both the seesaw model and Weinberg's model to describe Majorana neutrinos, attempts to assign a nontrivial charge conjugation transformation rule to each chiral fermion separately. But this common construction  is known to be operatorially ill-defined and, for example, the amplitude of the neutrinoless double beta decay vanishes if the vacuum is assumed to be invariant under the pseudo C-symmetry.

\end{abstract}
\section{Majorana as a superposition of chiral fermions}
The construction of the chiral fermion from the Dirac fermion is well-known. Also the construction of the Majorana fermion from the Dirac fermion is straightforward. In these cases, the charge conjugation operator is basically common between the Dirac fermion and the chiral fermion or the Dirac fermion and the Majorana fermion, respectively. But the construction of the Majorana fermion by a superposition of chiral fermions is less well understood.  

The Majorana fermion may be generally defined as a fermion which is identical to its anti-particle \cite{Majorana}, and it is usually defined as an eigenstate of charge conjugation symmetry $C$. This naive characterization becomes more involved when one constructs the Majorana fermion by a superposition of chiral fermions such as $\nu_{L} + C\overline{\nu_{L}}^{T}$, since the conventional charge conjugation is not defined for the left-handed  chiral fermion $\nu_{L}$ alone. We show that 
a more general definition of an anti-particle by the use of CP symmetry, which is a common practice in particle physics when C is ill-defined,  characterizes the Majorana neutrino constructed from chiral fermions in a natural manner and that the CP symmetry describes the entire physics of Majorana neutrinos in an extension of the Standard Model.  We also show that, if one attempts to specify deformed C and P separately by preserving the conventional CP for the Majorana neutrino formed  by a superposition of chiral fermions, such an attempt could lead to difficulties  depending on the choice of C and P.  The possible C operator other than the conventional one is constrained by the condition of the absence of Majorana-Weyl fermions in  $d=4$ dimensions.

\subsection{CP symmetry for a Majorana fermion}
  The  conventional C and P  are defined in textbooks on field theory \cite{Bjorken, Weinberg2} as the symmetries  of Dirac fermions with the convention $C=i\gamma^{2}\gamma^{0}$ and using $i\gamma^{0}$ for the parity \footnote{We define the parity of a Dirac fermion by $i\gamma^{0}$-parity, ${\cal P}\psi(x){\cal P}^{\dagger} = i\gamma^{0}\psi(t,-\vec{x})$, instead of the conventional $\gamma^{0}$-parity, since the Majorana fermion satisfying the classical relation $\psi(x)=C\overline{\psi(x)}^{T}$ is invariant under the parity thus defined $i\gamma^{0}\psi(t,-\vec{x}) = C\overline{i\gamma^{0}\psi(t,-\vec{x})}^{T}$. For consistency, we assign this convention of parity to charged leptons also, although this extra phase is cancelled in the lepton number conserving terms.  See \cite{Weinberg2} for the phase freedom of parity operation. See also \cite{Majorana, Racah}.},
\begin{eqnarray}\label{Dirac C and P}
&&{\cal C}\psi(x){\cal C}^{\dagger}=C\overline{\psi(x)}^{T},\ \ \ 
{\cal P}\psi(x){\cal P}^{\dagger}=i\gamma^{0}\psi(t,-\vec{x})
\end{eqnarray}  
which preserve the action invariant 
\begin{eqnarray}
\int d^{4}x {\cal L} = \int d^{4}x \{\overline{\psi}(x)[i\gamma^{\mu}\partial_{\mu} -m]\psi(x)\}.
\end{eqnarray}
The C and P symmetries translated to the chirally projected fields $\psi_{L,R}(x)$ are used for charged leptons and quarks in gauge theory and thus set the convention to analyze discrete symmetries in the Standard Model; the analyses of C, P and CP symmetries are performed using the chiral basis.  In the case of massive neutrinos, the transformation rules of the case of Dirac neutrinos, taken as a specific example,  give the definitions of C and P for chirally projected components~\footnote{We are not assuming that the massive neutrinos are actually Dirac fermions. We are illustrating that  these rules of C and P have deep mathematical and physical bases. } 
\begin{eqnarray}\label{conventional C and P}
&&{\cal C}\nu_{L}(x){\cal C}^{\dagger}=C\overline{\nu_{R}(x)}^{T},\ \ \ {\cal C}\nu_{R}(x){\cal C}^{\dagger}=C\overline{\nu_{L}(x)}^{T},\nonumber\\
&&{\cal P}\nu_{L}(x){\cal P}^{\dagger}=i\gamma^{0}\nu_{R}(t,-\vec{x}),\ \ \ {\cal P}\nu_{R}(x){\cal P}^{\dagger}=i\gamma^{0}\nu_{L}(t,-\vec{x}),\nonumber\\
&&({\cal P}{\cal C})\nu_{L}(x)({\cal P}{\cal C})^{\dagger}=i\gamma^{0}C\overline{\nu_{L}(t,-\vec{x})}^{T},\ \ \ ({\cal P}{\cal C})\nu_{R}(x)({\cal P}{\cal C})^{\dagger}=i\gamma^{0}C\overline{\nu_{R}(t,-\vec{x})}^{T}
\end{eqnarray}  
with
\begin{eqnarray}\label{chiral projection}
\nu_{R,L}(x)=(\frac{1\pm \gamma_{5}}{2})\nu(x).
\end{eqnarray}
These rules extracted from the Dirac fermion are mathematically consistent and the symmetries of the action 
\begin{eqnarray}
S &=& \int d^{4}x\{\overline{\nu_{L}}(x)i\gamma^{\mu}\partial_{\mu}\nu_{L}(x) + \overline{\nu_{R}}(x)i\gamma^{\mu}\partial_{\mu}\nu_{R}(x)\nonumber\\
&& - m\overline{\nu_{L}}(x)\nu_{R}(x)
- m\overline{\nu_{R}}(x)\nu_{L}(x)\}.
\end{eqnarray}
Physically, parity is defined as the mirror symmetry and one can check if the given Lagrangian is parity preserving or not using these rules. 
Good P naturally implies  left-right symmetry, and P is represented in the form of a doublet representation $\{\nu_{R}(x),\nu_{L}(x)\}$. The doublet representation of the charge conjugation C in \eqref{conventional C and P} is related to the absence of the Majorana-Weyl fermion in $d=4$ dimensions, which is a consequence of the representation of the Clifford algebra~\footnote{The substitution rule of charge conjugation symmetry $\nu_{L}(x) \rightarrow C\overline{\nu_{R}(x)}^{T}$, for example, is invariant when one multiplies the projection operator $(1-\gamma_{5})/2$ to both-hand sides, which is a basic requirement of the symmetry transformation law in field theory. On the other hand, $\nu_{R}(x)$ on the right-hand side arises from the fact that the charge conjugation reverses the signature of $\gamma_{5}$, which is related to the absence of the Majorana-Weyl fermion in $d=4$ dimensions.} ; intuitively, the absence of the Majorana-Weyl fermion is related to the fact that the charge conjugation inevitably changes the signature of 
$\gamma_{5}$ in $d=4$, namely, $\gamma_{5} \rightarrow -\gamma_{5}$. If one uses the charge conjugation operator other than \eqref{conventional C and P}, one always has a potential danger of spoiling the condition of the absence of Majorana-Weyl fermions in $d=4$. The CP transformation rules in \eqref{conventional C and P} are known to be good symmetries in rather general contexts.

The validity of general symmetry transformation rules \eqref{conventional C and P} are tested by the invariance of the given Lagrangian. For example, these rules \eqref{conventional C and P} imply that we assign only CP to the massless Weyl fermion 
\begin{eqnarray}
{\cal L}=\int d^{4}x \overline{\psi}_{L}(x)i\gamma^{\mu}\partial_{\mu}\psi_{L}(x)
\end{eqnarray}
without specifying C and P separately.

In the actual analyses of C, P and CP symmetries, the notion of field re-definition is important. The field re-definition in the present context  is a canonical transformation defined by linear transformations of field variables
which preserve the form of kinetic terms and thus   canonical anti-commutation relations.  The best known example of the canonical transformation  is the Kobayashi-Maskawa analysis of CP violation \cite{KM}.  A more general class of canonical transformation in the context of neutrino physics is known as the Pauli-Gursey  transformation \cite{Pauli, Gursey} and its generalization \cite{KF-PG}. The canonical transformation modifies the mass terms and interaction terms in general. The rule we adopt is that we apply the general discrete transformation rules in \eqref{conventional C and P} to {\em field variables defining the kinetic terms of the given Lagrangian after each canonical transformation}, since we understand that the canonical commutation relations characterize the Hilbert space. In this formulation, the C, P and CP symmetry transformations of field variables before the canonical transformation do not reproduce the C, P and CP transformations of field variables after the canonical transformation, respectively, in general \cite{KF-PG}, as is explicitly seen in the case of CP violating mass terms. More concrete examples of the canonical transformation shall be discussed later in connection with Weinberg's model of Majorana neutrinos \cite{Weinberg1} and also in connection with the Type-I or Type-I+II seesaw model \cite{Minkowski, Yanagida, Mohapatra}.

As for Majorana fermions, it is well-known that the Majorana fermions are defined in terms of the Dirac fermion $\psi_{D}(x)$ (in $d=4$ dimensions) 
\begin{eqnarray}\label{Majorana from Dirac}
\psi_{M\pm}(x)=\frac{1}{\sqrt{2}}[\psi_{D}(x) \pm C\overline{\psi_{D}(x)}^{T}]
\end{eqnarray} 
which satisfy the classical Majorana conditions $C\overline{\psi_{M\pm}(x)}^{T}=\pm \psi_{M\pm}(x)$ identically regardless of the choice of $\psi_{D}(x)$, i.e., $\psi_{M\pm}(x)$ are constrained variables. The operator definitions of the charge conjugation and parity are given using ${\cal C}$ and ${\cal P}$ defined for the Dirac fermion \eqref{Dirac C and P} by~\footnote{The definition $\psi_{M-}=\frac{1}{\sqrt{2}i}[\psi_{D}(x) - C\overline{\psi_{D}(x)}^{T}]$ with an imaginary factor $i$ which satisfies classical relation $\psi_{M-}=C\overline{\psi_{M-}}^{T}$ is often used, instead of our $\psi_{M-}(x)=\frac{1}{\sqrt{2}}[\psi_{D}(x) - C\overline{\psi_{D}(x)}^{T}]$, but this definition requires an anti-unitary ${\cal C}$ to maintain ${\cal C}\psi_{M-}{\cal C}^{\dagger}=\psi_{M-}$. }
\begin{eqnarray}\label{Conventional Majorana}
&&{\cal C}\psi_{M\pm}(x){\cal C}^{\dagger} =C\overline{\psi_{M\pm}(x)}^{T} =\pm \psi_{M\pm}(x),\nonumber\\
&&{\cal P}\psi_{M\pm}(x){\cal P}^{\dagger}=i\gamma^{0} \psi_{M\pm}(t,-\vec{x}).
\end{eqnarray}
The C symmetry of the Dirac fermion is represented in the form of a doublet representation $\{\psi_{M+}(x), \psi_{M-}(x)\}$ of Majorana fermions, but in a diagonalized form. The relations \eqref{Majorana from Dirac} show that  the Majorana fermion is defined on the same Hilbert space as the Dirac fermion with good C and P.
     
The common model of Majorana neutrinos such as Weinberg's model or the seesaw model  constructs Majorana neutrinos from chiral fermions. Weinberg's effective Lagrangian ~\cite{Weinberg1} constructs Majorana neutrinos from left-handed chiral fermions. There are many schemes of seesaw mechanism which construct Majorana fermions from chiral fermions. In the present paper we later discuss the so-called Type-I (or Type-I+II) model \cite{Minkowski, Yanagida, Mohapatra} in detail which contains the same number of left- and right-handed chiral fermions. The Type-II and Type-III models, which are accounted for in detail in the review article \cite{Xing} and monographs \cite{Fukugita, Giunti, Bilenky},  construct the Majorana fermions from left-handed chiral fermions after the mass diagonalization, and thus they may be included in the generic Weinberg-type models in the context of the present analysis.  
 
To be explicit, Weinberg's model of Majorana neutrinos~\cite{Weinberg1} is defined by an effective  Lagrangian 
\begin{eqnarray}\label{Weinberg-type}
{\cal L}&=&\overline{\nu_{L}}(x)i\delslash \nu_{L}(x)
-(1/2)\{\nu_{L}^{T}(x)Cm_{L}\nu_{L}(x) + h.c.\}
\end{eqnarray}
with a $3\times 3$ complex symmetric mass matrix $m_{L}$.
After the diagonalization of the symmetric complex mass matrix by the $3\times 3$ Autonne-Takagi factorization using a unitary $U$ \cite{Autonne-Takagi}
\begin{eqnarray}
U^{T}m_{L}U= M
\end{eqnarray}
with a real $3\times 3$ diagonal matrix $M$, we define 
\begin{eqnarray}\label{A-T transformation}
\nu_{L}(x)=U\tilde{\nu}_{L}(x)
\end{eqnarray}
and thus transfer the possible CP breaking to the PMNS mixing matrix which contains a mixing matrix coming from the charged lepton sector also in an extension of the Standard Model.
We then have (suppressing the tilde-symbol of $\tilde{\nu}_{L}(x)$)
\begin{eqnarray}\label{Weinberg-type2}
{\cal L}
&=&\overline{\nu_{L}}(x)i\delslash \nu_{L}(x)
-(1/2)\{\nu_{L}^{T}(x)CM\nu_{L}(x) + h.c.\}\nonumber\\
&=&(1/2)\{\overline{\psi}(x)i\delslash \psi(x)- \overline{\psi}(x)M\psi(x)\}          
\end{eqnarray}
where we defined
\begin{eqnarray}\label{Majorana in Weinberg}
\psi(x)\equiv \nu_{L}(x)+ C\overline{\nu_{L}}^{T}(x).
\end{eqnarray}
The field $\psi(x)$ satisfies the classical Majorana condition identically regardless of the choice of $\nu_{L}(x)$ (i.e., $\psi(x)$ is a constrained variable) 
\begin{eqnarray}\label{classical condition1}
\psi(x)=C\overline{\psi(x)}^{T}
\end{eqnarray}
and thus one may define
\begin{eqnarray}
{\cal C}_{M}\psi(x){\cal C}^{\dagger}_{M}=C\overline{\psi(x)}^{T}=\psi(x)
\end{eqnarray}
with a suitable ${\cal C}_{M}$ which we want to identify \footnote{The difference between the classical Majorana condition and the operatorial characterization is that the classical Majorana condition specifies only the field $\psi(x)$ as in \eqref{classical condition1}, while the operatorial definition requires a specification of transformation laws of its components $(\frac{1\pm \gamma_{5}}{2})\psi(x)$ also.}. 

In passing, we mention that the transformation \eqref{A-T transformation} provides an example of the canonical transformation we discussed above.  Both Lagrangians \eqref{Weinberg-type} and \eqref{Weinberg-type2}  have the same form of kinetic terms, and those kinetic terms indicate that we can apply the  CP symmetry \eqref{conventional C and P} to the field variables in these Lagrangians. The difference is that the mass terms in \eqref{Weinberg-type2} are invariant under CP transformation but the mass terms in \eqref{Weinberg-type} are not invariant in general.  Thus the CP transformation of $\tilde{\nu}_{L}(x)$ does not induce the CP transformation of $\nu(x)$ in general. The Autonne-Takagi transformation \eqref{A-T transformation} is not CP preserving. In the context of the Kobayashi-Maskawa analysis \cite{KM}, we regard \eqref{Weinberg-type2} as an analogue of the mass eigenstate, while \eqref{Weinberg-type} is an analogue of the original Yukawa-type mass matrix. In fact one can confirm that the propagator for $\tilde{\nu}_{L}(x)$ is diagonal (in flavor) in \eqref{Weinberg-type2}, while the propagator for $\nu_{L}(x)$ is not diagonal in \eqref{Weinberg-type} in general.  The possible CP breaking is analyzed by the PMNS matrix based on the fields $\tilde{\nu}_{L}(x)$ with well-defined asymptotic properties, and thus the entire CP breaking is attributed to weak interactions. We choose $\tilde{\nu}_{L}(x)$, which are denoted by $\nu_{L}(x)$ in \eqref{Weinberg-type2}, as primary variables to analyze physical implications in our analysis.

Coming back to the Majorana neutrinos, an interesting complication is that the doublet representations of C and P for the chirally projected components \eqref{conventional C and P} induced by the Dirac fermion do not reproduce the C and P symmetry transformation laws of the Majorana
fermion \eqref{Majorana in Weinberg}, of which transformation rules are also induced by the Dirac fermion,
\begin{eqnarray}
&&{\cal C}\psi(x){\cal C}^{\dagger}=\nu_{R}(x)+ C\overline{\nu_{R}}^{T}(x) \neq \psi(x),\nonumber\\
&&{\cal P}\psi(x){\cal P}^{\dagger}=i\gamma^{0}[\nu_{R}(t,-\vec{x})+ C\overline{\nu_{R}}^{T}(t,-\vec{x})] \neq i\gamma^{0} \psi(t,-\vec{x}).
\end{eqnarray}
Of course, this difficulty is related to the fact that we do not cover the left- and right-handed chiral fermion states symmetrically in \eqref{Weinberg-type2} and no $\nu_{R}$ is defined in the present model. But this relation implies that both the transition from the Dirac to chiral fermions and  the transition from the Dirac to Majorana fermions are straightforward, but the connection between the chiral and Majorana fermions is more subtle.

In the present case \eqref{Weinberg-type2},  C and P are not specified for the field $\nu_{L}$, but CP symmetry,
$({\cal P}{\cal C})\nu_{L}(x)({\cal P}{\cal C})^{\dagger}=i\gamma^{0}C\overline{\nu_{L}(t,-\vec{x})}^{T}$, in \eqref{conventional C and P}
is well-defined. We thus naturally characterize the Majorana fermion \eqref{Majorana in Weinberg} by CP symmetry
\begin{eqnarray}
&&({\cal P}{\cal C})[\nu_{L}(x)+ C\overline{\nu_{L}}^{T}(x)]({\cal P}{\cal C})^{\dagger}=i\gamma^{0}[ C\overline{\nu_{L}}^{T}(t,-\vec{x}) + \nu_{L}(t,-\vec{x})],
\end{eqnarray} 
namely,
\begin{eqnarray}\label{CP to define Majorana}
 ({\cal P}{\cal C})\psi(x)({\cal P}{\cal C})^{\dagger}= i\gamma^{0}C\overline{\psi(t,-\vec{x})}^{T}= i\gamma^{0}\psi(t,-\vec{x}).
\end{eqnarray}
The first equality in \eqref{CP to define Majorana} implies the operator relation while the second equality in \eqref{CP to define Majorana}  implies the classical Majorana condition \eqref{classical condition1} which holds identically in the sense that \eqref{classical condition1} holds irrespective of the choice of $\nu_{L}(x)$.  This characterization \eqref{CP to define Majorana} implies that we specify the Majorana neutrino with emphasis on $\nu_{L}(x)$.  The chiral fermion $\nu_{L}(x)$ appearing in $\psi(x)$, which is generated by a smooth renormalization group flow starting with the massless Weyl fermion in an extension of the Standard Model \cite{Weinberg1}, for example, has well-defined CP after mass diagonalization  just as the starting massless Weyl fermion.  The chiral component $\nu_{L}(x)$ of $\psi(x)$ describes  the weak interaction of the Standard Model 
\begin{eqnarray}\label{Weak-coupling}
\int d^{4}x[ (g/\sqrt{2})\bar{l}_{L}(x)\gamma^{\mu}W_{\mu}(x)U_{PMNS}\nu_{L}(x) + h.c.]
\end{eqnarray}
perfectly well, since the conventional C and P are broken in the chiral weak interaction and thus the specification of C and P for $\nu_{L}(x)$ in \eqref{Weinberg-type2} is not required. (An analysis of renormalization is somewhat involved in Weinberg's model of Majorana neutrinos.) In the present setting, we use the standard transformation rules of C, P and CP  for the chiral components of charged leptons and neutrinos in a uniform manner, although some of them are broken.

The analysis of CP is described by the PMNS matrix combined with the conventional CP symmetry properties of the charged lepton $l_{L}(x)$ and the chiral component $\nu_{L}(x)$ of the neutrino. The absence of the $U(1)$ phase freedom of $\nu_{L}(x)$ in \eqref{Weinberg-type2}, namely, the lepton number non-conservation,  is important to count an increase in the number of possible CP violating phases \cite{Takasugi}. 
The entire weak interaction is thus described by the chiral component  $\nu_{L}(x)$ using its CP property.
 The Majorana neutrinos characterized by CP symmetry retain certain information of their original chiral contents since the parity by itself is not specified in the present characterization of  Majorana neutrinos~\footnote{The use of CP symmetry (or more generally CPT) to define the physical Majorana neutrinos after the higher order corrections by C and P violating weak interactions has been considered in \cite{Kayser}. Our motivation is very different; the free Lagrangian 
 \eqref{Weinberg-type2} for chiral fermions is not invariant under the conventional C and P, which have clear mathematical and physical meanings,  but invariant under CP and thus we characterize the Majorana neutrinos by CP symmetry.}. 
 The fact that the free asymptotic field of the neutrino is a Majorana fermion  is assured by $\nu_{L}(x)=[(1-\gamma_{5})/2]\psi(x)$ and \eqref{CP to define Majorana} which contains the classical identity \eqref{classical condition1}. See also the discussion related to \eqref{Majorana basis2} below.

\subsection{Trivial C operator}
One may naturally want to describe the Majorana neutrino by a suitable quantum operator of C. Apparently, a discontinuous deformation of discrete symmetry operators from the ones in \eqref{conventional C and P} is required to describe the chiral fermion $\nu_{L}(x)$ by such an operator C. One may think of the  
 deformed symmetry operators \cite{KF-PG} \footnote{This deformation may be compared to the pseudo C-symmetry in subsection 1.3. Both retain the original CP symmetry but C and P separately are very different.} 
 \begin{eqnarray}\label{Majorana basis}
 {\cal C}_{M}=1, \ \ \ {\cal P}_{M}={\cal C}{\cal P}
 \end{eqnarray}
which  consist of well-defined operators $1$ and ${\cal C}{\cal P}$ in \eqref{Weinberg-type2}; this  shows that the vacuum of the chiral fermion and the vacuum of the Majorana fermion share the same CP symmetry. The choice \eqref{Majorana basis} defines the Majorana fermion $\psi(x)$ in \eqref{Majorana in Weinberg}
 \begin{eqnarray}\label{Majorana basis2}
{\cal C}_{M}\psi(x){\cal C}_{M}^{\dagger}=\psi(x)
, \ \ \
{\cal P}_{M}\psi(x){\cal P}_{M}^{\dagger}=i\gamma^{0}\psi(t,-\vec{x})
\end{eqnarray}
with ${\cal C}_{M}{\cal P}_{M}={\cal C}{\cal P}$. Literally taken, the first relation in \eqref{Majorana basis2} is satisfied by any functional form of $\psi(x)$, and in this sense only the CP operation contained in the second relation supplemented by the identity 
\eqref{classical condition1} is substantial. Since only the  CP  operator acts effectively, this choice \eqref{Majorana basis2} may be regarded as a subset of the characterization of the Majorana by CP in \eqref{CP to define Majorana}; the charge conjugation operator ${\cal C}_{M}$ gives rise to  ${\cal C}_{M}\psi(x){\cal C}_{M}^{\dagger} = \psi(x) = C\overline{\psi(x)}^{T}$ when applied to the Majorana fermion, but this is physically equivalent to the classical Majorana condition $\psi(x) = C\overline{\psi(x)}^{T}$ which is used in \eqref{CP to define Majorana}.
  
The formulation \eqref{Majorana basis} leads to a formal enhancement of discrete symmetries in  \eqref{Weinberg-type2} by assigning C and P to the chiral component $\nu_{L}(x)=(\frac{1-\gamma_{5}}{2})\psi(x)$,
\begin{eqnarray} \label{Majorana basis3} 
{\cal C}_{M}\nu_{L}(x){\cal C}_{M}^{\dagger}=\nu_{L}(x) =C\overline{\nu_{R}}^{T}(x), \ \ 
{\cal P}_{M}\nu_{L}(x){\cal P}_{M}^{\dagger}=i\gamma^{0}C\overline{\nu_{L}(t,-\vec{x})}^{T}=i\gamma^{0}\nu_{R}(t,-\vec{x})
\end{eqnarray}
by defining a variable $\nu_{R}(x)
\equiv (\frac{1+\gamma_{5}}{2})\psi(x) = C\overline{\nu_{L}}^{T}(x)$, and 
\begin{eqnarray}\label{Majorana basis4}
{\cal C}_{M}\nu_{R}(x){\cal C}_{M}^{\dagger}=C\overline{\nu_{L}}^{T}(x), \ \  {\cal P}_{M}\nu_{R}(x){\cal P}_{M}^{\dagger}=i\gamma^{0}\nu_{L}(t,-\vec{x}).
\end{eqnarray}
These transformation rules are mathematically consistent and imply perfect left-right symmetry  expected for a Majorana fermion $\psi=\nu_{L}+\nu_{R}$.

These rules \eqref{Majorana basis3} and \eqref{Majorana basis4} may be compared to the rules in \eqref{conventional C and P}. The physical degrees of freedom of a Majorana fermion $\psi=\nu_{L}+\nu_{R}$ are the same as either a left-handed chiral fermion or a right-handed chiral fermion but not both, contrary to the case of a Dirac fermion in \eqref{conventional C and P} where the left- and right-handed components are independent.  If one measures the right-handed projection of the Majorana fermion $\psi$, for example, one obtains the chiral freedom $\nu_{R}$ and simultaneously the information of $\nu_{L}$ also. 

The deformation \eqref{Majorana basis} is a specific choice of the definition of C-symmetry with emphasis on the generated Majorana fermion by preserving CP symmetry.  The trivial C may be natural from a point of view of the Majorana fermion defined in an abstract manner, but one may keep in mind that the fermion 
$\psi(x) = \nu_{R}(x)+ C\overline{\nu_{R}}^{T}(x)$
which has a completely different meaning from \eqref{Majorana in Weinberg} in the context of chiral  gauge theory, also defines a left-right symmetric Majorana fermion in the present formulation.
The ``Majorana-Weyl condition'' \eqref{Majorana basis3}
\begin{eqnarray}
{\cal C}_{M}\nu_{L}(x){\cal C}_{M}^{\dagger}=\nu_{L}(x)
\end{eqnarray}
in $d=4$ does not lead to a mathematical contradiction since ${\cal C}_{M}$ is trivial, besides $\nu_{L}(x)=C\overline{\nu_{R}(x)}^{T}$ by definition. The operators \eqref{Majorana basis} imply an assignment of $C_{M}$, $P_{M}$ and $C_{M}P_{M}$ to a massless Weyl fermion in the vanishing mass limit in \eqref{Weinberg-type2} and thus the equivalence of the massless Majorana neutrino and the massless Weyl neutrino, and both become Majorana fermions. This is a disturbing aspect of the use of the present charge conjugation operator \cite{KF-PG}.

The choice \eqref{Majorana basis} shows that one can define a consistent operator ${\cal C}_{M}$ which defines a Majorana fermion constructed from a chiral fermion. But the physical advantage of the use of this operator ${\cal C}_{M}$ for  
$\psi=\nu_{L} + C\overline{\nu_{L}}^{T}$ in SM is not very obvious; for example, it does not directly prohibit  the $U(1)$ phase change $\nu_{L}(x)\rightarrow e^{i\alpha}\nu_{L}(x)$ in the analysis of CP in the weak interaction \eqref{Weak-coupling} \cite{Takasugi} since the Majorana condition ${\cal C}_{M}e^{i\alpha}\nu_{L}(x){\cal C}_{M}=e^{i\alpha}\nu_{L}(x)$  or $\psi(x)= e^{i\alpha}\nu_{L}(x)+ e^{-i\alpha}C\overline{\nu_{L}}^{T}(x) = C\overline{\psi(x)}^{T}$ allows it. 
In comparison, the simpler characterization by CP in  \eqref{CP to define Majorana} describes the entire weak interactions including the counting of CP breaking phases without referring to C, as already explained. (The Lagrangian \eqref{Weinberg-type2} has no U(1) symmetry in terms of $\nu_{L}(x)$ but it has the symmetry in terms of $\psi(x)$ in the sense $\psi(x)= e^{i\alpha}\nu_{L}(x)+ e^{-i\alpha}C\overline{\nu_{L}}^{T}(x) = C\overline{\psi(x)}^{T}$.)

\subsection{Pseudo C-symmetry}  
One may still want to define a nontrivial charge conjugation rule to each chiral component separately and  define the Majorana fermion in a more direct manner.
The ``pseudo C-symmetry'' (this naming was suggested in \cite{FT1} to distinguish it from the conventional C-symmetry
\eqref{conventional C and P}) was invented as a result of such efforts. This scheme, which is commonly used to define Majorana neutrinos, in particular, in the seesaw model, thus attempts to assign a nontrivial charge conjugation rule to each chiral component separately. To be precise, one starts with the definition of ``pseudo C-symmetry'' $\tilde{{\rm C}}$ defined by the substitution rules (including $\nu_{R}$ to prepare for the analysis of the Type-I or Type-I+II seesaw model later) \cite{Xing, Fukugita, Giunti, Bilenky}
\begin{eqnarray}\label{pseudo C}
\tilde{{\rm C}}: \ \ \nu_{L}(x) \rightarrow C\overline{\nu_{L}(x)}^{T}, \ \ \  \nu_{R}(x) \rightarrow C\overline{\nu_{R}(x)}^{T}.
\end{eqnarray}
One then has for the classical Majorana field \eqref{Majorana in Weinberg} by noting $ C\overline{\nu_{L}(x)}^{T}\rightarrow \nu_{L}(x)$ by the above rule \eqref{pseudo C},
\begin{eqnarray}
\tilde{{\rm C}}: \ \ \psi(x)\rightarrow C\overline{\nu_{L}(x)}^{T}+\nu_{L}(x)= \psi(x)
\end{eqnarray}
namely, the Majorana condition is maintained.  The CP symmetry for the Majorana fermion is then defined by the composition rule
\begin{eqnarray}
\tilde{{\rm C}}\tilde{{\rm P}}:\ \ \psi(x)=[C\overline{\nu_{L}(x)}^{T}+\nu_{L}(x)]\rightarrow  i\gamma^{0}[C\overline{\nu_{L}(t,-\vec{x})}^{T}+\nu_{L}(t,-\vec{x})] = i\gamma^{0}\psi(t,-\vec{x})
\end{eqnarray}
if one defines  the ``pseudo P-symmetry'' $\tilde{{\rm P}}$ by
\begin{eqnarray}\label{pseudo P}
\tilde{{\rm P}}:\ \ \nu_{L}(x)\rightarrow i\gamma^{0}\nu_{L}(t,-\vec{x}), \ \ \  \nu_{R}(x)\rightarrow i\gamma^{0}\nu_{R}(t,-\vec{x})
\end{eqnarray}
for each chiral component separately.
The CP symmetry of chiral fermions is then defined by 
\begin{eqnarray}
\tilde{{\rm C}}\tilde{{\rm P}}:\ \ \nu_{L}(x) \rightarrow i\gamma^{0}C\overline{\nu_{L}(t,-\vec{x})}^{T}, \ \ \  \nu_{R}(x)\rightarrow i\gamma^{0}C\overline{\nu_{R}(t,-\vec{x})}^{T}
\end{eqnarray}
which agree with the conventional definition of CP symmetry 
\eqref{conventional C and P}. One can also confirm that the action defined by 
\eqref{Weinberg-type2} is formally invariant under $\tilde{{\rm C}}$ and $\tilde{{\rm P}}$, separately.

It thus appears that the pseudo C-symmetry and the pseudo P-symmetry which satisfy $\tilde{{\rm C}}\tilde{{\rm P}}=CP$ can define the Majorana neutrinos consistently. 
But these symmetries, which look natural by assigning the common vacuum to Majorana fermions and  chiral fermions,  are ill-defined operatorially when carefully examined. If one assumes the existence of {\em unitary operators} which generate these symmetries, 
the pseudo C-symmetry $\tilde{{\cal C}}\nu_{L}(x)\tilde{{\cal C}}^{\dagger}=C\overline{\nu_{L}}^{T}$ gives a natural solution of the relation $\tilde{{\cal C}}\psi(x)\tilde{{\cal C}}^{\dagger}               =C\overline{\psi(x)}^{T}$, namely, 
\begin{eqnarray}\label{Pseudo C2}
\tilde{{\cal C}}\nu_{L}(x)\tilde{{\cal C}}^{\dagger}+ \tilde{{\cal C}}C\overline{\nu_{L}(x)}^{T}\tilde{{\cal C}}^{\dagger} = C\overline{\nu_{L}(x)}^{T}+ \nu_{L}(x)
\end{eqnarray}
and similarly, the pseudo P-symmetry  $\tilde{{\cal P}}\nu_{L}(x)\tilde{{\cal P}}^{\dagger} = i\gamma^{0}\nu_{L}(t,-\vec{x})$ is a natural solution of $\tilde{{\cal P}}\psi(x)\tilde{{\cal P}}^{\dagger} = i\gamma^{0}\psi(t, -\vec{x})$, namely,
\begin{eqnarray}\label{Pseudo P2}
\tilde{{\cal P}}\nu_{L}(x)\tilde{{\cal P}}^{\dagger}+ \tilde{{\cal P}}C\overline{\nu_{L}(x)}^{T}\tilde{{\cal P}}^{\dagger}
=i\gamma^{0}\nu_{L}(t,-\vec{x}) + i\gamma^{0}C\overline{\nu_{L}(t,-\vec{x})}^{T}
\end{eqnarray}
with $\tilde{{\cal C}}\tilde{{\cal P}}={\cal C}{\cal P}$, although $\tilde{{\cal P}}\nu_{L}(x)\tilde{{\cal P}}^{\dagger} = i\gamma^{0}\nu_{L}(t,-\vec{x})$ spoils the idea of parity as a mirror symmetry. 

  These pseudo-symmetry operations, however, are known to lead to disturbing results
by noting $\nu_{L}(x)=(\frac{1-\gamma_{5})}{2})\nu_{L}(x)$ \cite{FT1,FT2,FT3}
\begin{eqnarray}\label{operatorial inconsistency}
&& \tilde{{\cal C}}\nu_{L}(x)\tilde{{\cal C}}^{\dagger} = (\frac{1-\gamma_{5})}{2})\tilde{{\cal C}}\nu_{L}(x)\tilde{{\cal C}}^{\dagger}=(\frac{1-\gamma_{5})}{2})C\overline{\nu_{L}(x)}^{T} =0,
\end{eqnarray}
and also
\begin{eqnarray}\label{operatorial inconsistency2}
&&\tilde{{\cal P}}\nu_{L}(x)\tilde{{\cal P}}^{\dagger} =(\frac{1-\gamma_{5})}{2})\tilde{{\cal P}}\nu_{L}(x)\tilde{{\cal P}}^{\dagger}=(\frac{1-\gamma_{5})}{2}) i\gamma^{0}\nu_{L}(t,-\vec{x})=0   
\end{eqnarray}
since both $C\overline{\nu_{L}(x)}^{T}$ and $i\gamma^{0}\nu_{L}(t,-\vec{x})$ are right-handed. As an alternative to \eqref{operatorial inconsistency}, one may start with $(\frac{1-\gamma_{5}}{2})\psi(x)=\nu_{L}(x)$ and obtain
$\tilde{{\cal C}}(\frac{1-\gamma_{5}}{2})\psi(x)\tilde{{\cal C}}^{\dagger}
=(\frac{1-\gamma_{5}}{2})\tilde{{\cal C}}\psi(x)\tilde{{\cal C}}^{\dagger}
=(\frac{1-\gamma_{5}}{2})\psi(x)$ and $\tilde{{\cal C}}\nu_{L}(x)\tilde{{\cal C}}^{\dagger}=C\overline{\nu_{L}(x)}^{T}=(\frac{1+\gamma_{5}}{2})\psi(x)$,
namely, $(\frac{1-\gamma_{5}}{2})\psi(x)=(\frac{1+\gamma_{5}}{2})\psi(x)$
which implies \footnote{The relation $(\frac{1-\gamma_{5}}{2})\psi(x)=(\frac{1+\gamma_{5}}{2})\psi(x)$ means $\nu_{L}(x)=C\overline{\nu_{L}(x)}^{T}$ which is understood as the Majorana-Weyl condition on $\nu_{L}(x)$ in $d=4$ dimensions and thus no solution. This property is related to the failure of chirality conservation requirement by the pseudo C-symmetry \cite{FT1}.} 
\begin{eqnarray}\label{operatorial inconsistency3}
(\frac{1-\gamma_{5}}{2})\psi(x)=(\frac{1+\gamma_{5}}{2})\psi(x)=0.
\end{eqnarray}

In the level of substitution rules also, these symmetry operations are ill-defined.
This is seen by considering the free action of the Majorana fermion  using 
$\psi(x)=\nu_{L}+ C\overline{\nu_{L}}^{T}$,
\begin{eqnarray}\label{free-Majorana}
S_{\rm Majorana}&=&\frac{1}{2}\int d^{4}x\,\overline{\psi(x)}[i\delslash -M]\psi(x)\nonumber\\
&=&\int d^{4}x \Big\{ \overline{\nu_{L}}i\delslash \nu_{L} -\frac{1}{2}\nu_{L}^{T}CM\nu_{L} - \frac{1}{2}\overline{\nu_{L}}MC\overline{\nu_{L}}^{T} \Big\}\nonumber\\
&=&\int d^{4}x \Big\{\overline{\nu_{L}}i\delslash \frac{(1-\gamma_{5})}{2}\nu_{L}(x)-\frac{1}{2}\nu_{L}^{T}CM\frac{(1-\gamma_{5})}{2}\nu_{L} + h.c. \Big\}.
\end{eqnarray}
We emphasize that these three expressions are identical.
If one assumes the transformation rule of pseudo C-symmetry,
$\nu_{L}(x)\rightarrow  C\overline{\nu_{L}(x)}^{T}$, as in  \eqref{pseudo C}, it turns out that  the first and second expressions in \eqref{free-Majorana} are invariant under the transformation, while the last expression leads to a vanishing Lagrangian \cite{FT1}. 
Similarly, one may assume a transformation rule of pseudo P-symmetry, $\nu_{L}(x)\rightarrow  i\gamma^{0}\nu_{L}(t,-\vec{x})$,
as in \eqref{pseudo P}, then  the first and second expressions in \eqref{free-Majorana} are invariant under the transformation, while the last expression leads to a vanishing Lagrangian.  This implies that one cannot decide if the pseudo C-symmetry and pseudo P-symmetry are good symmetries of \eqref{free-Majorana}; one cannot decide if the fermion $\psi(x)$ defined in \eqref{free-Majorana} is a Majorana fermion when one uses the pseudo C-symmetry and the pseudo P-symmetry.

We note that the puzzling aspects in \eqref{free-Majorana} arise from the substitution rules \eqref{pseudo C} and \eqref{pseudo P}, irrespective of the existence or non-existence of the quantum  operators $\tilde{{\cal C}}$ and $\tilde{{\cal P}}$ (although in the framework of field theory, we assume the operator representations of valid substitution rules). Consequently, the example \eqref{free-Majorana} shows that even as substitution rules, the pseudo C-symmetry and pseudo P-symmetry are ill-defined. 
In comparison, $\tilde{{\cal C}}\tilde{{\cal P}}$, which agrees with the conventional CP without referring to $\tilde{{\cal C}}$ and $\tilde{{\cal P}}$ separately, is consistent in every respect. 

It has been shown in \cite{FT1} that the pseudo C-symmetry is formally obtained from a truncation of the CP symmetry
\begin{eqnarray}\label{truncation of CP}
CP:\ \nu_{L}(x) \rightarrow i\gamma^{o}C\overline{\nu_{L}(t,-\vec{x})}^{T}\
\Rightarrow \ \tilde{C}:\ \nu_{L}(x) \rightarrow C\overline{\nu_{L}(t,\vec{x})}^{T}
\end{eqnarray}
in any CP invariant theory. Namely, the pseudo C-symmetry is obtained from CP symmetry by removing the pre-factor $i\gamma^{o}$ and restoring the spatial inversion $-\vec{x} \rightarrow \vec{x}$, and it is still formally a symmetry of the CP invariant theory such as \eqref{Weinberg-type2}. But the pseudo C-symmetry is operatorially ill-defined if one examines the transformation more carefully as in \eqref{free-Majorana} \footnote{The pseudo C-symmetry is thus related to the use of CP symmetry we suggest in the present paper; the difference is that $\tilde{{\cal C}}$ fails to act consistently on the component $(\frac{1-\gamma_{5}}{2})\psi(x)$ of $\psi(x)$, $\tilde{{\cal C}}(\frac{1-\gamma_{5}}{2})\psi(x)\tilde{{\cal C}}^{\dagger}=\tilde{{\cal C}}(\frac{1-\gamma_{5}}{2})\nu_{L}(x)\tilde{{\cal C}}^{\dagger}=0$ as in \eqref{operatorial inconsistency} which is related to the failure of the chirality conservation \cite{FT1}, while the CP symmetry is operatorially consistent for both $\psi(x)$ and $(\frac{1-\gamma_{5}}{2})\psi(x)$.}. 

So far we have emphasized the disturbing aspects of the pseudo C-symmetry based on the statements in \eqref{operatorial inconsistency}, \eqref{operatorial inconsistency2} and \eqref{free-Majorana} which are mathematical facts.  On the other hand, the pseudo C-symmetry has been used in the past in many papers on Majorana neutrinos in the seesaw model, for example, without any obvious contradictions. One may thus want to understand the basic reason of the apparent phenomenological success of the formulation with the pseudo C-symmetry.
The pseudo C-symmetry $\tilde{{\cal C}}\nu_{L}(x)\tilde{{\cal C}}^{\dagger}=C\overline{\nu_{L}}^{T}$ as a natural solution of the relation 
\begin{eqnarray}\label{Pseudo C2}
\tilde{{\cal C}}\psi(x)\tilde{{\cal C}}^{\dagger}               =\tilde{{\cal C}}\nu_{L}(x)\tilde{{\cal C}}^{\dagger}+ \tilde{{\cal C}}C\overline{\nu_{L}(x)}^{T}\tilde{{\cal C}}^{\dagger} = \psi(x)
\end{eqnarray}
is an interesting finding.  We have also shown the existence of a pseudo P-symmetry $\tilde{{\cal P}}\nu_{L}(x)\tilde{{\cal P}}^{\dagger}=i\gamma^{0}\nu_{L}(t, -\vec{x})$ as a solution of
\begin{eqnarray}\label{Pseudo P2}
\tilde{{\cal P}}\psi(x)\tilde{{\cal P}}^{\dagger}               =\tilde{{\cal P}}\nu_{L}(x)\tilde{{\cal P}}^{\dagger}+ \tilde{{\cal P}}C\overline{\nu_{L}(x)}^{T}\tilde{{\cal P}}^{\dagger}= i\gamma^{0}\psi(t, -\vec{x})
\end{eqnarray}
which satisfies the condition $\tilde{{\cal C}}\tilde{{\cal P}}={\cal C}{\cal P}$. If one uses the pseudo C-symmetry only for the purpose of the identification of a Majorana neutrino \eqref{Pseudo C2} (and the pseudo P-symmetry only in the form \eqref{Pseudo P2}),  and if one uses the conventional CP symmetry to analyze the weak interactions, one can analyze the weak interaction phenomenology successfully without encountering any contradictions. 
We have emphasized in \eqref{CP to define Majorana} that the proper use of CP symmetry describes all the physics aspects of Majorana neutrinos in an extension of SM successfully.

In conclusion of this subsection, the secret of the practical success of the formulation with the pseudo C-symmetry is that people used the pseudo C-symmetry only to identify the Majorana fermion \eqref{Pseudo C2} and simply did not use the problematic aspects of the pseudo C-symmetry (and the pseudo P-symmetry) to analyze weak interactions, which can be described well by the conventional CP symmetry without referring to the pseudo C-symmetry.  A problematic aspect of the pseudo C-symmetry, if one should use it directly in weak interaction phenomenology, shall be illustrated in Section 3.

\section{Majorana neutrino from Dirac-type fermions}

\subsection{Type-I Seesaw model}
We have already mentioned that Majorana fermions are defined in terms of the Dirac fermion $\psi_{D}(x)$,
$\psi_{M\pm}(x)=\frac{1}{\sqrt{2}}[\psi_{D}(x) \pm C\overline{\psi_{D}(x)}^{T}]$, as in in  \eqref{Majorana from Dirac}. 
The operator definitions of the charge conjugation and parity are naturally given using nontrivial ${\cal C}$ and ${\cal P}$ defined for the Dirac fermion \eqref{Dirac C and P} in the manner \eqref{Conventional Majorana}, since $\psi_{D}(x)$ and $\psi_{M\pm}(x)$ are defined on the same left-right symmetric state space.
This option is not available for the Majorana fermions defined by chiral fermions in  Weinberg's model.  But we shall show that this option is available in the seesaw model of Type-I or Type-I+II, which contains the same number of left- and right-handed chiral fermions, if one uses the generalized Pauli-Gursey transformation which is a canonical transformation. We thus have more options, the construction from Dirac fermions as in  \eqref{Conventional Majorana} and the use of CP symmetry \eqref{CP to define Majorana},  to define Majorana neutrinos in the seesaw model of Type-I or Type-I+II, in addition to the use of trivial C \eqref{Majorana basis} which may be regarded as a subset of \eqref{CP to define Majorana}. This analysis is useful to illustrate the fact that the vacua for the Majorana neutrino and the chiral neutrino after mass diagonalization are identical from the point of view of CP symmetry but they are very different from the point of view of charge conjugation symmetry; these different vacua are, however,  smoothly connected by a canonical transformation. 

We first recapitulate the basic aspects of the seesaw model. The seesaw model for the three generations of neutrinos starts with
\begin{eqnarray}\label{Lagrangian}
{\cal L}&=&\overline{\nu}_{L}(x)i\gamma^{\mu}\partial_{\mu}\nu_{L}(x)+\overline{\nu}_{R}(x)i\gamma^{\mu}\partial_{\mu}\nu_{R}(x)\nonumber\\
&-&\overline{\nu}_{L}(x)m_{D} \nu_{R}(x)
-(1/2)\nu_{L}^{T}(x)C m_{L}\nu_{L}(x)\nonumber\\
&-&(1/2)\nu_{R}^{T}(x)C m_{R}\nu_{R}(x) + h.c.,
\end{eqnarray}
where $m_{D}$ is a complex $3\times 3$ Dirac mass matrix, and $m_{L}$ and $m_{R}$ are $3\times 3$ complex Majorana-type matrices. The anti-symmetry of the matrix $C$
 and Fermi statistics imply that $m_{L}$ and $m_{R}$ are symmetric. This is the Lagrangian of neutrinos with Dirac and Majorana mass terms. For $m_L=0$, it represents the classical seesaw Lagrangian of Type-I. In the following, we shall call the expression \eqref{Lagrangian}, which may be properly called a Type-I+II seesaw model \cite{Xing, Pontecorvo}, as the seesaw Lagrangian.
 
We start with writing the mass term as 
\begin{eqnarray}
(-2){\cal L}_{mass}=
\left(\begin{array}{cc}
            \overline{\nu_{R}}&\overline{\nu_{R}^{C}}
            \end{array}\right)
\left(\begin{array}{cc}
            m_{R}& m_{D}\\
            m^{T}_{D}&m_{L}
            \end{array}\right)
            \left(\begin{array}{c}
            \nu_{L}^{C}\\
            \nu_{L}
            \end{array}\right) +h.c.,
\end{eqnarray}
where 
\begin{eqnarray}
\nu_{L}^{C}\equiv C\overline{\nu_{R}}^T, \ \ \ \nu_{R}^{C}\equiv C\overline{\nu_{L}}^T.  
\end{eqnarray}
Note that $\nu_{L}^{C}$ and $\nu_{R}^{C}$ are left-handed and right-handed, respectively. 
Since the mass matrix appearing is complex and symmetric, we can diagonalize it 
by a $6 \times 6$ unitary transformation $U$ (Autonne-Takagi factorization \cite{Autonne-Takagi}) as
\begin{eqnarray}\label{mass diagonalization}
            U^{T}
            \left(\begin{array}{cc}
            m_{R}& m_{D}\\
            m_{D}& m_{L}
            \end{array}\right)
            U
            =\left(\begin{array}{cc}
            M_{1}&0\\
            0&-M_{2}
            \end{array}\right)    ,        
\end{eqnarray}
where  $M_{1}$ and $M_{2}$ are $3\times 3$ real diagonal matrices (characteristic values). We choose one of the eigenvalues as $-M_{2}$ instead of $M_{2}$ since it is a natural choice in the case of a single generation model.

We thus have
\begin{eqnarray}\label{exact-mass}
(-2){\cal L}_{mass}
&=& \left(\begin{array}{cc}
            \overline{\tilde{\nu}_{R}}&\overline{\tilde{\nu}_{R}^{C}}
            \end{array}\right)
\left(\begin{array}{cc}
            M_{1}&0\\
            0&-M_{2} 
            \end{array}\right)            
            \left(\begin{array}{c}
            \tilde{\nu}_{L}^{C}\\
            \tilde{\nu}_{L}
            \end{array}\right) +h.c.,                       
\end{eqnarray}
with
\begin{eqnarray} \label{variable-change}          
            &&\left(\begin{array}{c}
            \nu_{L}^{C}\\
            \nu_{L}
            \end{array}\right)
            = U \left(\begin{array}{c}
            \tilde{\nu}_{L}^{C}\\
            \tilde{\nu}_{L}
            \end{array}\right)
           ,\ \ \ \ 
            \left(\begin{array}{c}
            \nu_{R}\\
            \nu_{R}^{C}
            \end{array}\right)
            = U^{\star} 
            \left(\begin{array}{c}
            \tilde{\nu}_{R}\\
            \tilde{\nu}_{R}^{C}
            \end{array}\right).          
\end{eqnarray}
Hence we can write 
\begin{eqnarray}\label{exact-solution}
{\cal L}
&=&(1/2)\{\overline{\tilde{\nu}_{L}}(x)i\delslash\tilde{\nu}_{L}(x)+ \overline{\tilde{\nu}_{L}^{C}}(x)i\delslash \tilde{\nu}_{L}^{C}(x)+\overline{\tilde{\nu}_{R}}(x)i\delslash \tilde{\nu}_{R}(x)\nonumber\\
&& \ \ \ \ \ + \overline{\tilde{\nu}_{R}^{C}}(x)i\delslash \tilde{\nu}_{R}^{C}(x)\}\nonumber\\
&-&(1/2)\left(\begin{array}{cc}
            \overline{\tilde{\nu}_{R}},&\overline{\tilde{\nu}_{R}^{C}}
            \end{array}\right)
\left(\begin{array}{cc}
            M_{1}&0\\
            0&-M_{2} 
            \end{array}\right)            
            \left(\begin{array}{c}
            \tilde{\nu}_{L}^{C}\\
            \tilde{\nu}_{L}
            \end{array}\right) +h.c..
\end{eqnarray}
In the present transformation \eqref{variable-change} in terms of a unitary matrix,  one can confirm that the conditions of canonical transformation
\begin{eqnarray} \tilde{\nu}_{L}^{C}=C\overline{\tilde{\nu}_{R}}^{T}, 
\ \ \
\tilde{\nu}_{R}^{C}=C\overline{\tilde{\nu}_{L}}^{T}
\end{eqnarray}
hold after the transformation.  See \eqref{basic-condition} below. This canonical transformation with a unitary $U$ has been identified as a special case of the generalized Pauli-Gursey transformation \cite{KF-PG}. Phenomenologically, this unitary transformation transfers the possible CP breaking in the neutrino sector to the PMNS weak mixing matrix in the seesaw model. 

The Lagrangian
\eqref{Lagrangian} is then written in the form (by suppressing the tilde symbol for the chiral states $\tilde{\nu}_{R,L}$ which diagonalize the mass terms)
\begin{eqnarray}\label{exact-solution2}
{\cal L}
&=&(1/2)\{\overline{\psi_{+}}(x)i\delslash \psi_{+}(x)+ \overline{\psi_{-}}(x)i\delslash \psi_{-}(x)\}\nonumber\\
&-&(1/2)\{\overline{\psi_{+}}M_{1}\psi_{+}+ \overline{\psi_{-}}M_{2}\psi_{-}\}            
\end{eqnarray}
where 
\begin{eqnarray}\label{Majorana-variables3}
\psi_{+}(x)=\nu_{R}+ C\overline{\nu_{R}}^{T}, \ \ 
\psi_{-}(x)=\nu_{L}- C\overline{\nu_{L}}^{T}
\end{eqnarray}
which satisfy the classical Majorana conditions identically (i.e., $\psi_{\pm}(x)$ are constrained variables)
\begin{eqnarray}\label{Majorana-variables4}
C\overline{\psi_{+}(x)}^{T}= \psi_{+}(x), \ \ 
C\overline{\psi_{-}(x)}^{T}= -\psi_{-}(x).
\end{eqnarray}
But the charge conjugation operation defined for the chirally projected Dirac fermion \eqref{conventional C and P} does not work
\begin{eqnarray}
&&{\cal C}\psi_{+}(x){\cal C}^{\dagger}=\nu_{L}+ C\overline{\nu_{L}}^{T}\neq \psi_{+}(x),\nonumber\\
&&{\cal C}\psi_{-}(x){\cal C}^{\dagger}=-\nu_{R}+ C\overline{\nu_{R}}^{T}\neq - \psi_{-}(x).
\end{eqnarray}
Our suggestion is thus to characterize the Majorana neutrinos using CP, namely, using the CP transformation laws of chiral fermions 
\eqref{conventional C and P} which are the good symmetry of \eqref{exact-solution}, 
\begin{eqnarray}
&&({\cal P}{\cal C})\psi_{+}(x)({\cal P}{\cal C})^{\dagger}=i\gamma^{0}C\overline{\psi_{+}(t,-\vec{x})}^{T}=i\gamma^{0}\psi_{+}(t,-\vec{x}), \nonumber\\
&&({\cal P}{\cal C})\psi_{-}(x)({\cal P}{\cal C})^{\dagger}=i\gamma^{0}C\overline{\psi_{-}(t,-\vec{x})}^{T}=-i\gamma^{0}\psi_{-}(t,-\vec{x})
\end{eqnarray} 
that are consistent in every respect and describe all the physics aspects of the seesaw model, as explained already

In contrast, in most of the common treatments of the seesaw model \cite{Xing, Fukugita, Giunti, Bilenky}, one adopts the ``pseudo C-symmetry'' \eqref{pseudo C} (and ``pseudo P-symmetry'' \eqref{pseudo P}, although not often mentioned), 
which formally appear to work (using the operator notation)
\begin{eqnarray}\label{pseudo C-transformation}          
&&\tilde{{\cal C}}\psi_{+}(x)\tilde{{\cal C}}^{\dagger}=C\overline{\nu_{R}(x)}^{T}+\nu_{R}(x)= \psi_{+}(x),\nonumber\\
&&\tilde{{\cal C}}\psi_{-}(x)\tilde{{\cal C}}^{\dagger}=C\overline{\nu_{L}(x)}^{T}-\nu_{L}(x)=-\psi_{-}(x),\nonumber\\
&&\tilde{{\cal P}}\psi_{+}(x)\tilde{{\cal P}}^{\dagger}=i\gamma^{0}[\nu_{R}(t,-\vec{x})+C\overline{\nu_{R}(t,-\vec{x})}^{T}]= i\gamma^{0}\psi_{+}(t,-\vec{x}),\nonumber\\
&&\tilde{{\cal P}}\psi_{-}(x)\tilde{{\cal P}}^{\dagger}=i\gamma^{0}[\nu_{L}(t,-\vec{x})-C\overline{\nu_{L}(t,-\vec{x})}^{T}]=i\gamma^{0}\psi_{-}(t,-\vec{x})
\end{eqnarray}
by assigning both the charge conjugation and parity transformation rules to Majorana fermions.
One can also confirm that the actions constructed from \eqref{exact-solution} and \eqref{exact-solution2} are formally invariant under $\tilde{{\cal C}}$ and $\tilde{{\cal P}}$. However, one encounters operatorial ill-definedness for these symmetry transformation rules in both quantum and substitution-rule levels when carefully examined, as already analyzed in \eqref{operatorial inconsistency}, \eqref{operatorial inconsistency2} and \eqref{free-Majorana}, respectively.

\subsection{Generalized Pauli-Gursey transformation}

A way to resolve the complications associated with the definition of Majorana neutrinos using the C symmetry in the seesaw model of Type-I or Type-I+II, which are briefly summarized above, has been discussed in detail using a relativistic analogue of the Bogoliubov transformation \cite{FT2, FT3, Bogoliubov}. The generalized Pauli-Gursey transformation, which is closely related to the Bogoliubov transformation,  is more transparent in the treatment of  CP symmetry \cite{KF-PG}.  The generalized Pauli-Gursey transformation is defined  by the transformation \eqref{variable-change} but now with {\em arbitrary} $U(6)$ \cite{KF-PG}
\begin{eqnarray}\label{Pauli-Gursey}        
            &&\left(\begin{array}{c}
            \nu_{L}^{C}\\
            \nu_{L}
            \end{array}\right)
            = U \left(\begin{array}{c}
            \tilde{\nu}_{L}^{C}\\
            \tilde{\nu}_{L}
            \end{array}\right)
           ,\ \ \ \ 
            \left(\begin{array}{c}
            \nu_{R}\\
            \nu_{R}^{C}
            \end{array}\right)
            = U^{\star} 
            \left(\begin{array}{c}
            \tilde{\nu}_{R}\\
            \tilde{\nu}_{R}^{C}
            \end{array}\right)          
\end{eqnarray}
which still satisfies the conditions of canonical transformation (namely, the anti-commutation relations are preserved after the transformation) \footnote{The fundamental condition \eqref{basic-condition}, which is essential to define a canonical transformation,
is satisfied after the change of variables \eqref{Pauli-Gursey}, if one notes 
$\tilde{\nu}_{L}=(U^{\dagger})_{21}\nu_{L}^{C} + (U^{\dagger})_{22}\nu_{L}$ and $\tilde{\nu}_{R}^{C}=(U^{\dagger})^{\star}_{21}\nu_{R}+(U^{\dagger})^{\star}_{22}\nu_{R}^{C}$ using $3\times 3$ submatrices defined by
\begin{eqnarray}
U^{\dagger}=\left(\begin{array}{cc}
            (U^{\dagger})_{11}&(U^{\dagger})_{12}\\
            (U^{\dagger})_{21}&(U^{\dagger})_{22} 
            \end{array}\right).\nonumber
\end{eqnarray}
}
\begin{eqnarray}\label{basic-condition} \tilde{\nu}_{L}^{C}=C\overline{\tilde{\nu}_{R}}^{T}, 
\ \ \
\tilde{\nu}_{R}^{C}=C\overline{\tilde{\nu}_{L}}^{T}.
\end{eqnarray}
Recall that $\tilde{\nu}_{L}^{C}$ and $\tilde{\nu}_{R}^{C}$ in our definition are left-handed and right-handed, respectively.
The  generalized Pauli-Gursey transformation with an arbitrary unitary transformation $U(6)$ in \eqref{Pauli-Gursey}  mixes fermions and anti-fermions, and thus changes the definition of the vacuum together with C and  P symmetries defined on each vacuum.  Historically, the Pauli-Gursey transformation was defined for a single generation with $U(2)$ \cite{Pauli, Gursey}.  See also \cite{Schechter, Balantekin}
for related analyses.

A general strategy is then to choose a suitable ``Pauli frame'' defined by the generalized Pauli-Gursey transformation, analogously to the ``Lorentz frame'' in the terminology of Lorentz transformation, such that the seesaw Lagrangian is expressed in terms of Dirac-type variables only, which  then allows us to define the Majorana neutrinos in a natural manner using the conventional C and P symmetries \eqref{Dirac C and P}.  
For this purpose, we consider a further $6\times 6$ real generalized Pauli-Gursey transformation $O(6)$ (which is of course  included in $U(6)$) in addition to \eqref{exact-solution}, that is orthogonal and thus preserves CP \cite{KF-PG}, by 
\begin{eqnarray} \label{orthogonal-variable-change}          
            &&\left(\begin{array}{c}
            \tilde{\nu}_{L}^{C}\\
            \tilde{\nu}_{L}           
            \end{array}\right)
            = O \left(\begin{array}{c}
            N_{L}^{C}\\
            N_{L}
            \end{array}\right)
           ,\ \ \ \ 
            \left(\begin{array}{c}
           \tilde{\nu}_{R}\\
            \tilde{\nu}_{R}^{C}
            \end{array}\right)
            = O
            \left(\begin{array}{c}
            N_{R}\\
            N_{R}^{C}
            \end{array}\right).         
\end{eqnarray}
The exact solution \eqref{exact-solution} is then rewritten, by choosing a specific $6\times6$ orthogonal transformation
\begin{eqnarray}\label{orthogonal1}
O=\frac{1}{\sqrt{2}}\left(\begin{array}{cc}
            1&1\\
            -1&1
            \end{array}\right)
\end{eqnarray}
where $1$ stands for a $3\times3$ unit matrix,
in the form 
\begin{eqnarray}\label{BCS-like}
{\cal L}
&=&(1/2)\{\overline{N}(x)i\delslash N(x)+ \overline{N^{C}}(x)i\delslash N^{C}(x)\}\nonumber\\
&-&(1/4)\{\overline{N}(M_{1}+M_{2})N+ \overline{N^{C}}(M_{1}+M_{2})N^{C}\}\nonumber\\
&-&(1/4)[\overline{N}(M_{1}-M_{2})N^{C}+\overline{N^{C}}(M_{1}-M_{2})N].
\end{eqnarray}
This Lagrangian is expressed in terms of Dirac-type variables $N(x)$ and $N^{C}(x)$ only (noting that $N^{C}_{L}(x)=C\overline{N_{R}}^{T}(x)$ and $N^{C}_{R}(x)=C\overline{N_{L}}^{T}(x)$, respectively) and invariant under the conventional C, P and CP defined by 
\begin{eqnarray}\label{C-P-CP-of-N}
C&:& N(x) \leftrightarrow N^{C}(x)=C\overline{N}^{T}(x),\nonumber\\
P&:& N(x) \rightarrow i\gamma^{0}N(t,-\vec{x}), \ \ N^{C}(x) \rightarrow i\gamma^{0}N^{C}(t,-\vec{x}),\nonumber\\
CP&:& N(x) \rightarrow i\gamma^{0}N^{C}(t,-\vec{x}), \ \ N^{C}(x) \rightarrow i\gamma^{0}N(t,-\vec{x}).
\end{eqnarray}
The weak mixing angles are now determined by $U^{\prime}=UO$ instead of $U$ in \eqref{mass diagonalization}.
This is also the essence of a relativistic analogue of the Bogoliubov transformation~\cite{FT2,FT3}~\footnote{It is interesting that the notion of multiple vacua was considered at about the same time  independently by Pauli and by BCS and Bogoliubov around 1957.}.
It is significant  that only the Dirac-type particles $N(x)$ and $N^{C}(x)$ with the conventional C, P and CP transformation properties appear in this specific Pauli frame where the Lagrangian \eqref{BCS-like} becomes C and P invariant and the chiral structure disappears, although the lepton number is violated.

We now make a renaming of variables 
\begin{eqnarray} \label{Majorana-variables}          
            \psi_{+}(x)=\frac{1}{\sqrt{2}}(N(x)+N^{C}(x)), \ \ 
            \psi_{-}(x)=\frac{1}{\sqrt{2}}(N(x)-N^{C}(x)),           
\end{eqnarray}
which satisfies the classical Majorana conditions identically, $C\overline{\psi_{+}(x)}^{T}=\psi_{+}(x)$ and $C\overline{\psi_{-}(x)}^{T}=-\psi_{-}(x)$,
and we obtain
\begin{eqnarray}\label{Majorana1}
{\cal L}
&=&(1/2)\{\overline{\psi_{+}}(x)i\delslash \psi_{+}(x)+ \overline{\psi_{-}}(x)i\delslash \psi_{-}(x)\}\nonumber\\
&-&(1/2)\{\overline{\psi_{+}}M_{1}\psi_{+}+ \overline{\psi_{-}}M_{2}\psi_{-}\}.
\end{eqnarray}
After this renaming of variables, we find the transformation laws of $\psi_{\pm}(x)$ induced by those of $N$ and $N^{C}$ in \eqref{C-P-CP-of-N},
\begin{eqnarray}\label{final-C-P-CP}
C&:& \psi_{+}(x) \rightarrow \psi_{+}(x), \ \ \psi_{-}(x) \rightarrow - \psi_{-}(x),\nonumber\\
P&:& \psi_{+} \rightarrow i\gamma^{0}\psi_{+}(t,-\vec{x}), \ \ \psi_{-}(x) \rightarrow i\gamma^{0}\psi_{-}(t,-\vec{x}),\nonumber\\
CP&:& \psi_{+}(x) \rightarrow i\gamma^{0}\psi_{+}(t,-\vec{x}), \ \ \psi_{-}(x) \rightarrow -i\gamma^{0}\psi_{-}(t,-\vec{x})
\end{eqnarray}
which naturally keep the Lagrangian \eqref{Majorana1} invariant.
 When one defines a nontrivial unitary charge conjugation operator 
 \begin{eqnarray}
{\cal C}_{N}N(x){\cal C}^{\dagger}_{N}=N^{C}(x)=C\overline{N}^{T}(x),
\end{eqnarray}
the operator ${\cal C}_{M}={\cal C}_{N}$ gives rise to 
\begin{eqnarray} 
{\cal C}_{M}\psi_{\pm}(x){\cal C}^{\dagger}_{M}=\pm \psi_{\pm}(x)
\end{eqnarray} 
which is an analogue of the conventional definition of the Majorana fermion in terms of a Dirac fermion in \eqref{Majorana from Dirac} and \eqref{Conventional Majorana} (in the actual operator construction, it is easier to construct ${\cal C}_{M}$ first).  We can also define parity consistently for Majorana fermions
\begin{eqnarray}
{\cal P}_{M}\psi_{\pm}(x){\cal P}^{\dagger}_{M}=i\gamma^{0}\psi_{\pm}(t,-\vec{x})
\end{eqnarray}
if one defines ${\cal P}_{M}={\cal P}_{N}$ with 
${\cal P}_{N}N(x){\cal P}^{\dagger}_{N}=i\gamma^{0}N(t,-\vec{x})$ and thus ${\cal P}_{N}N^{C}(x){\cal P}^{\dagger}_{N}=i\gamma^{0}N^{C}(t,-\vec{x})$.
We thus determine 6 Majorana fermions $\psi_{\pm}(x)$ (each contains 3 flavor freedom) in the conventional manner after a suitable choice of generalized Pauli-Gursey transformation as above.

The essence of the generalized Pauli-Gursey transformation is that we mix fermions and anti-fermions as in \eqref{Pauli-Gursey} and thus leading to the existence of multiple vacua.
This change of the vacuum allows us to define the Majorana neutrinos in the seesaw model of Type-I or Type-I+II consistently using the standard definitions of C and P for Dirac-type fermions on a suitably defined new vacuum.  
The fact that the orthogonal transformation $O(6)$, which preserves CP but modifies C and P \cite{KF-PG}, works in \eqref{BCS-like} shows that the chiral fermions and the Majorana fermions share the same CP symmetry in the real diagonal mass bases; in the presence of left- and right-components, the chiral fermions are re-arranged to be Dirac-type fermions with this CP kept in tact.

\section{Neutrinoless double beta decay}

One may wonder if the formal analyses in the present paper have any physical implications in neutrino phenomenology. The two basic physical properties implied by the possible Majorana neutrinos are the neutrinoless double beta decay and an increase in the number of CP violating phases in the PMNS matrix compared to the CKM matrix.  The modified CP phase freedom has been mentioned in connection with \eqref{Weak-coupling}. We here discuss  the neutrinoless double beta decay to show the physical relevance of the formal analyses in the present paper.  The neutrinoless double beta decay is described by the weak interaction Lagrangian  \cite{Doi}
\begin{eqnarray}\label{beta-decay}
\int d^{4}x {\cal L}_{\rm Weak}&=&\int d^{4}x[ (g/\sqrt{2})\bar{l}_{L}(x)\gamma^{\mu}W_{\mu}(x)U_{PMNS}\nu_{L}(x) + h.c.]\nonumber\\
&=&\int d^{4}x[ (g/\sqrt{2})\bar{l}_{L}(x)\gamma^{\mu}W_{\mu}(x)U_{PMNS}\frac{(1-\gamma_{5})}{2}\psi_{M}(x) + h.c.]
\end{eqnarray}
with the three generations of charged leptons $l_{L}(x)$ and the PMNS mixing matrix $U_{PMNS}$ in the case of Weinberg's model $\psi_{M}(x)=\nu_{L}(x)+C\overline{\nu_{L}}^{T}(x)$ in  \eqref{Majorana in Weinberg}. A similar analysis is valid for the seesaw model when expressed in terms of chiral fermions as in \eqref{Majorana-variables3}.
A necessary condition of the neutrinoless double beta decay is that not all  the time-ordered correlations of the neutrino mass eigenstates 
\begin{eqnarray}\label{propagator}
&&\langle 0|T^{\star} \nu_{L}(x)\nu_{L}(y)|0\rangle = \langle 0|T^{\star} \frac{(1-\gamma_{5})}{2}\psi_{M}(x)\frac{(1-\gamma_{5})}{2}\psi_{M}(y)|0\rangle
\end{eqnarray}
vanish  in the second order perturbation in $ {\cal L}_{\rm Weak}$~\cite{Doi}.  It is interesting that the neutrinoless double beta decay is neatly characterized by the vacuum expectation value of the T-product of neutrino fields without referring to charged leptons.
We suppress the hadronic sector.

If a unitary operator $\tilde{{\cal C}}$ which generates the pseudo C-symmetry  \eqref{pseudo C} exists and if the (neutrino) vacuum $|0\rangle$ should be invariant 
\begin{eqnarray}
\tilde{{\cal C}}^{\dagger}|0\rangle=|0\rangle 
\end{eqnarray}
and thus $\langle 0|\tilde{{\cal C}}=\langle0|$, one can prove that all of the above correlations vanish
\begin{eqnarray}\label{no double beta decay}
\langle 0|T^{\star} \nu_{L}(x)\nu_{L}(y)|0\rangle&=&\langle 0|T^{\star} [(\frac{1-\gamma_{5}}{2})\nu_{L}](x)\nu_{L}(y)|0\rangle\nonumber\\
&=&\langle 0|\tilde{{\cal C}}T^{\star} [(\frac{1-\gamma_{5}}{2})\nu_{L}](x)\nu_{L}(y)\tilde{{\cal C}}^{\dagger}|0\rangle\nonumber\\
&=&\langle 0|T^{\star} [(\frac{1-\gamma_{5}}{2})C\overline{\nu_{L}}^{T}](x)[C\overline{\nu_{L}}^{T}](y)|0\rangle\nonumber\\
&=&0
\end{eqnarray}
where we used $\nu_{L}(x)=(\frac{1-\gamma_{5}}{2})\nu_{L}(x)$ and $\tilde{{\cal C}}\nu_{L}(x)\tilde{{\cal C}}^{\dagger} = C\overline{\nu_{L}(x)}^{T}$  and the fact that  $C\overline{\nu_{L}}^{T}(x)$ is right-handed. Namely, no neutrinoless double beta decay would take place in the second order of perturbation in weak interactions. 
The same conclusion holds if one assumes that the substitution rule $\nu_{L}(x)\rightarrow C\overline{\nu_{L}}^{T}(x)$ in \eqref{pseudo C} is a good symmetry of the action of the neutrino sector \eqref{Weinberg-type2} \footnote{Alternatively, by recalling that the pseudo C-symmetry implies $\frac{(1-\gamma_{5})}{2}\psi_{M}(x)\rightarrow C\overline{\frac{(1-\gamma_{5})}{2}\psi_{M}(x)}^{T}=\frac{(1+\gamma_{5})}{2}\psi_{M}(x)$, one may conclude $\langle 0|T^{\star} \frac{(1-\gamma_{5})}{2}\psi_{M}(x)\frac{(1-\gamma_{5})}{2}\psi_{M}(y)|0\rangle = \\
\langle 0|T^{\star} \frac{(1-\gamma_{5})}{2}[\frac{(1-\gamma_{5})}{2}\psi_{M}(x)][\frac{(1-\gamma_{5})}{2}\psi_{M}(y)]|0\rangle\rightarrow 0$, which is consistent with \eqref{operatorial inconsistency3}.}. The conclusion \eqref{no double beta decay} is a consequence of the operatorially ill-defined pseudo C-symmetry \eqref{operatorial inconsistency}, and it illustrates a problematic aspect of the pseudo C-symmetry when it is used directly in weak interaction phenomenology except for the identification of the Majorana neutrino.
In contrast, one can confirm that CP invariance $({\cal P}{\cal C})^{\dagger}|0\rangle=|0\rangle$ of the vacuum with $({\cal P}{\cal C})\nu_{L}(x)({\cal P}{\cal C})^{\dagger}=i\gamma^{0}C\overline{\nu_{L}(t,-\vec{x})}^{T}$ as in \eqref{CP to define Majorana} is consistent and leads to the neutrinoless double beta decay in general, as the explicit evaluation of \eqref{propagator} indicates. 

One can also confirm that other choices of C symmetry operators such as 
 C invariance ${\cal C}_{N}^{\dagger}|0\rangle=|0\rangle$ of the vacuum after the generalized Pauli-Gursey transformation in \eqref{final-C-P-CP} with ${\cal C}_{N}\nu_{L\pm}(x){\cal C}^{\dagger}_{N}=\pm \nu_{L\pm}(x)$ where 
\begin{eqnarray}
\nu_{L\pm}(x)\equiv \frac{1}{\sqrt{2}}(N_{L}(x)\pm C\overline{N_{R}}^{T}(x)) = (\frac{1-\gamma_{5}}{2})\psi_{\pm}(x),
\end{eqnarray}
  does not lead to any constraint and gives the conventional result for the above correlation of neutrino fields.  Note that one of $\nu_{L\pm}(x)$ with a smaller mass is physically relevant in the seesaw model.  Also, the trivial ${\cal C}$ in \eqref{Majorana basis}, which may be regarded as a subset of \eqref{CP to define Majorana}, does not give rise to complications in the analysis of \eqref{propagator}.

\section{Summary and conclusion} 
In the Standard Model where the neutrinos are treated as massless Weyl fermions, one  describes all the fermions including quarks and charged leptons by the conventional C, P and CP symmetries \eqref{conventional C and P}. The proposal of the present paper is to characterize Majorana neutrinos by the CP symmetry using the same definitions of the discrete symmetries \eqref{conventional C and P} even after one adds the lepton number violating Majorana-type mass terms to neutrinos.  This formulation has been illustrated for the effective Lagrangian of Majorana neutrinos suggested by Weinberg, where the Majorana-type mass terms are generated by the process of the renormalization group flow~\cite{Weinberg1}.
    
We here restate the historical backgrounds leading to the above proposal. 
Weinberg's effective Lagrangian after the mass diagonalization assumes the form 
\begin{eqnarray}\label{Weinberg-type2-C}
{\cal L}
&=&\overline{\nu_{L}}(x)i\delslash \nu_{L}(x)
-(1/2)\{\nu_{L}^{T}(x)CM\nu_{L}(x) + h.c.\}\nonumber\\
&=&(1/2)\{\overline{\psi}(x)i\delslash \psi(x)- \overline{\psi}(x)M\psi(x)\}          
\end{eqnarray}
where we defined
\begin{eqnarray}\label{Majorana in Weinberg-C}
\psi(x)\equiv \nu_{L}(x)+ C\overline{\nu_{L}}^{T}(x).
\end{eqnarray}
Similarly, another class of Majorana neutrinos in the form $\psi(x)= \nu_{R}(x) + C\overline{\nu_{R}(x)}^{T}$ generally appear in the seesaw model after the diagonalization of Majorana masses, but for the moment we concentrate on \eqref{Majorana in Weinberg-C}. The neutrinos thus constructed satisfy the classical Majorana condition identically independent of the choice of $\nu_{L}(x)$
\begin{eqnarray}\label{Classical Majorana-C}
\psi(x) = C\overline{\psi(x)}^{T}
\end{eqnarray}
and in this operation, $\nu_{L}(x)$ and $C\overline{\nu_{L}(x)}^{T}$ are interchanged. It is thus very tempting to identify a charge conjugation operation by
\begin{eqnarray}
\tilde{{\cal C}}\nu_{L}(x) \tilde{{\cal C}}^{\dagger} = C\overline{\nu_{L}(x)}^{T}, \ \ \tilde{{\cal C}}C\overline{\nu_{L}(x)}^{T} \tilde{{\cal C}}^{\dagger} = \nu_{L}(x)
\end{eqnarray}
which ensures $\tilde{{\cal C}}\psi(x) \tilde{{\cal C}}^{\dagger} = C\overline{\psi(x)}^{T} = \psi(x)$ in \eqref{Classical Majorana-C}. This operation denoted by $\tilde{{\cal C}}$ has been named ``pseudo C-symmetry'' in \cite{FT1}; it has been shown that the pseudo C-symmetry is operatorially inconsistent and thus very different from any sensible definition of charge conjugation \cite{FT1,FT2}. In the present paper, we also defined ``pseudo P-symmetry'' by 
\begin{eqnarray}
\tilde{{\cal P}}\nu_{L}(x)\tilde{{\cal P}}^{\dagger} = i\gamma^{0}\nu_{L}(t,-\vec{x}), \ \ \tilde{{\cal P}}C\overline{\nu_{L}(x)}^{T} \tilde{{\cal P}}^{\dagger} = i\gamma^{0}C\overline{\nu_{L}(t,-\vec{x})}^{T}
\end{eqnarray}
which satisfies $\tilde{{\cal P}}\psi(x)\tilde{{\cal P}}^{\dagger} = i\gamma^{0}\psi(t,-\vec{x})$ and reproduces the conventional CP symmetry $\tilde{{\cal C}}\tilde{{\cal P}}={\cal C}{\cal P}$. From this definition of $\tilde{{\cal P}}$, which does not reverse the helicity under the parity operation, it is obvious that the pseudo P-symmetry does not represent the physical mirror symmetry and mathematically  inconsistent $\tilde{{\cal P}}\nu_{L}(x) \tilde{{\cal P}}^{\dagger} = (\frac{1-\gamma_{5}}{2})\tilde{{\cal P}}\nu_{L}(x) \tilde{{\cal P}}^{\dagger} = (\frac{1-\gamma_{5}}{2}) i\gamma^{0}\nu_{L}(t,-\vec{x}) =0$.

One might still argue that the pseudo C-symmetry has been used in many papers on the Majorana neutrinos successfully in the past without any obvious contradictions and thus the above mathematical inconsistencies may be physically irrelevant technical details. In section 3 of the present paper, we demonstrated that the faithful use of pseudo C-symmetry leads to the vanishing amplitude of the neutrinoless double-beta decay, which is the most fundamental manifestation of Majorana neutrinos,  and thus our formal analysis is physically very relevant. We also explained in some detail why people who used pseudo C-symmetry in the past have not directly encountered mathematical inconsistencies in subsection 1.3. 

In the specific cases of Type-I or Type-I+II seesaw models where the same number of left- and right-handed chiral fermions appear, it has been shown that one can formulate Majorana neutrinos using the conventional C symmetry defined for Dirac-type fermions by exploiting the similarity of the seesaw Lagrangian with the BCS Lagrangian of superconductivity and the associated  Bogoliubov transformation~\cite{FT1, FT2}. In these specific cases, one can also use a generalization of the idea of Pauli and Gursey~\cite{KF-PG} and arrive at the same conclusion as the use of the relativistic analogue of the Bogoliubov transformation. These facts are summarized in section 2 for the sake of completeness. 
Technically, the notion of the canonical transformation in field theory among the vacua with different C-symmetry plays a central role in these reformulations of the seesaw models.      

Coming back to Weinberg's model of Majorana neutrinos or Type-II or Type-III seesaw models, where only the left-handed chiral fermions appear, the above idea of the Bogoliubov transformation or the Pauli-Gursey transformation does not work in these models.
We thus need a more universal scheme which is applicable to a more general class of models. In the present paper we have proposed to use the CP symmetry which is a good symmetry of Weinberg's model expressed in terms of chiral fermions   after the mass diagonalization \eqref{Weinberg-type2-C}.
Under the conventional C symmetry \eqref{conventional C and P}, we have $\nu_{L}(x) \rightarrow C\overline{\nu_{R}(x)}^{T}$ and thus C is not a good symmetry in \eqref{Weinberg-type2-C} just as in the case of a Weyl fermion, but CP is a good symmetry of the Lagrangian \eqref{Weinberg-type2-C}. We thus characterize the Majorana fermion $\psi(x) = \nu_{L}(x)+ C\overline{\nu_{L}}^{T}(x)$ by 
\begin{eqnarray}\label{CP to define Majorana-C}
 ({\cal P}{\cal C})\psi(x)({\cal P}{\cal C})^{\dagger}= i\gamma^{0}C\overline{\psi(t,-\vec{x})}^{T}= i\gamma^{0}\psi(t,-\vec{x}).
\end{eqnarray}
The first equality in \eqref{CP to define Majorana-C} implies the operator relation while the second equality in \eqref{CP to define Majorana-C}  implies the classical Majorana condition \eqref{Classical Majorana-C} which holds identically in the sense that \eqref{Classical Majorana-C} holds irrespective of the choice of $\nu_{L}(x)$.  (This characterization of the Majorana neutrinos  after the mass diagonalization \eqref{CP to define Majorana-C} also works for the two classes of Majorana neutrinos $\nu_{L}(x)+ C\overline{\nu_{L}}^{T}(x)$ and $\nu_{R}(x)+ C\overline{\nu_{R}}^{T}(x)$ appearing in the Typr-I or Type-I+II seesaw
models.)

   The characterization \eqref{CP to define Majorana-C}  implies that we specify the Majorana neutrino with emphasis on $\nu_{L}(x)$ and it is advantageous when one analyzes weak interactions which are expressed by $\nu_{L}(x)$. All the weak interactions of leptons including CP symmetry are analyzed in a manner identical to the sector of quarks. The only difference is that we have no lepton number $U(1)$ symmetry for $\nu_{L}(x)$ in \eqref{Weinberg-type2-C} and thus the number of CP violating phases is generally increased compared to the quark sector.
   
An unconventional aspect of the present proposal is that we do not define Majorana neutrinos by a charge conjugation {\em operator}. The fact that the neutrinos are Majorana fermions are expressed by the classical Majorana condition contained in the second relation of \eqref{CP to define Majorana-C}.  Alternatively, if one wishes, one could formally define the {\em deformed} symmetry generators ${\cal C}_{M}=1$ and ${\cal P}_{M} = {\cal C}{\cal P}$ \cite{KF-PG} to characterize the Majorana neutrinos by the operator ${\cal C}_{M}\psi(x){\cal C}_{M}^{\dagger} = \psi(x)$ and ${\cal C}_{M}\nu_{L}(x){\cal C}_{M}^{\dagger} = {\cal C}_{M}(\frac{1-\gamma_{5}}{2})\psi(x){\cal C}_{M}^{\dagger} = \nu_{L}(x)$ as was discussed in subsection 1.2, but these deformed symmetries are shown to have little practical advantages in the analyses of weak interactions. Physically, the classical Majorana condition $C\overline{\psi(x)}^{T}=\psi(x)$, which is used in \eqref{CP to define Majorana-C}, carries the same information as the trivial operator ${\cal C}_{M}\psi(x){\cal C}_{M}^{\dagger}= \psi(x) = C\overline{\psi(x)}^{T}$ applied to the Majorana neutrino $\psi(x)$.  

  In conclusion, we have shown that we can describe all the physics aspects of possible Majorana neutrinos in  an extension of the Standard Model using the CP symmetry based on the conventional C and P symmetries in \eqref{conventional C and P} extracted from Dirac fermions. 
\\

I thank A. Tureanu for helpful discussions on  Majorana neutrinos. I also thank T. Maskawa for an informative conversation on the Pauli-Gursey transformation and J. Arafune for clarifying the phenomenological implications of $i\gamma^{0}$-parity.  The present work is supported in part by JSPS KAKENHI (Grant No.18K03633).

\end{document}